 \newlength\smallfigwidth
 \documentclass[aps,apl,twocolumn,floatfix,showpacs,amsmath,amssymb]{revtex4-1}

\usepackage{graphicx}
\usepackage[hypertex]{hyperref}
\def\ba{\begin{eqnarray}}
\def\ea{\end{eqnarray}}
\def\be{\begin{equation}}
\def\ee{\end{equation}}

\begin{document}

\preprint{UFV}

\title{Bound-states and polarized charged zero modes in three-dimensional topological insulators induced by a magnetic vortex}

\author{J.M. Fonseca}
\email{jakson.fonseca@ufv.br}
\affiliation{Grupo de F\'{\i}sica/ICET/CRP, Universidade Federal de Vi\c cosa,
Rodovia MG-230 Km 7, Cep 38810-000, Rio Parana\'iba, Minas Gerais, Brazil.}
\author{W.A. Moura-Melo}
\email{winder@ufv.br\\ URL: https://sites.google.com/site/wamouramelo/home}
\author{A.R. Pereira}
\email{apereira@ufv.br\\ URL: https://sites.google.com/site/quantumafra/}
 \affiliation{Departamento de F\'isica,
Universidade Federal de
Vi\c cosa, Vi\c cosa, 36570-000, Minas Gerais, Brazil \\
}

\date{\today}

\begin{abstract}
By coating a three-dimensional topological insulator (TI) with a ferromagnetic film supporting an in-plane magnetic vortex, one breaks the time-reversal symmetry (TRS) without generating a mass gap. It rather yields electronic states bound to the vortex center which have different probabilities associated with each spin mode. In addition, its associate current (around the vortex center) is partially polarized with an energy gap separating the most excited bound state from the scattered ones. Charged zero-modes also appear as fully polarized modes localized near the vortex center. From the magnetic point of view, the observation of such a special current in a TI-magnet sandwich comes about as an alternative technique for detecting magnetic vortices in magnetic thin films.
\end{abstract}
\pacs{73.20.-r, 75.30.Hx, 85.75.-d}

%
%73.43.-f quantum hall effects
%78.20.Ls: Magneto-optical effects
%03.65.Vf: Phases: geometric; dynamic or topological
%73.20.-r: Electron states at surfaces and interfaces
%75.30.Hx: Magnetic impurity interactions
%85.75.-d: Magnetoelectronics; spintronics:
%04.20.-q: Classical general relativity
%

\maketitle

%-------------------------------------------------------------------
\section{Introduction}
%-------------------------------------------------------------------
Topological Insulators (TI's) are materials that exhibit a bulk insulating behavior able to conduct boundary states protected by time-reversal symmetry (TRS) \cite{TI review}. There is a nontrivial number that classifies a TI, distinguishing it from a trivial insulator (like the integer quantum Hall effect). The topological number is associated with the Bloch wave-function that describes the bulk electrons in the momentum space and it can assume two values: $\nu_0=0$ (or even) specifies an ordinary insulator while $\nu_0=1$ (or odd) accounts for a topological insulator. In the quantum Hall effect, the external magnetic field breaks TRS and leads to nontrivial topological numbers for the edge electrons. In the case of a TI, the spin-orbit coupling is the physical origin of the nontrivial topological number but preserving TRS. At present, such a behavior has been observed in $Bi_{x}Sb_{1-x}$ alloys,
$Bi_{2}Se_{3}$, $Bi_{2}Te_{3}$, $Sb_{2}Se_{3}$ crystals \cite{Hsieh 2008, Chen, Xia} and
in ternary intermetallic Heusler compounds \cite{Chadov, Wray}.
The reason for the considerable interest in topological insulators lies in their potential
for producing new physical phenomena as well as future technological applications. These rely on the special properties of the topologically protected electronic surface states, composed of an odd number of Dirac fermions, familiar to the physics of graphene \cite{castro-neto-review}. When the topological number changes, like in the interface between a TI and vacuum (or any other ordinary insulator) the gap vanishes giving rise to gapless states in the TI surface, which are topologically protected by bulk properties of the Bloch wave-function. In other words, the physical manifestation of the topological order comes about in the form of protected gapless surface states that are robust against damaging the surface by chemical or mechanical means such as
alterations in its shape or orientation with respect to the crystal lattice, or even by disordering the bulk,
as long as such changes are applied in moderation \cite{TI review}.\\

The above statements apply to systems that respect TRS, that is, nonmagnetic TI in zero external magnetic field. When TRS is broken, even by a weak magnetic perturbation, a gap can be open up into the spectrum of the topologically protected surface states.
When a gap opens as a consequence of a magnetic perturbation, the resulting surface is not an ordinary
insulator; instead, it is a quantum Hall insulator with properties similar to those of the familiar
quantum Hall systems realized in two-dimensional electron gas \cite{y. l. chen}. A tunable
energy gap at the surface Dirac point provides a means to control the surface electric transport, which
is of great importance for applications.

Topological order in proximity to magnetism has been considered as one of the main topics in the field
\cite{y. l. chen,liu-prl, Essim, Qi, garate and franz 2010, nomura and nagaosa 2010}. In principle, it can
be realized by several means, such as the application of weak magnetic fields through a Zeeman term\cite{Q10}
$H_{Z}=-g\mu_{B}\vec{s}\cdot\vec{B}$, or through exchange coupling to magnetic thin films\cite{Qi}
$H_{int}=-\Delta\vec{M}\cdot\vec{s}$, and magnetic impurities\cite{W10,Liu09,y. l. chen}
$H_{imp}=\sum_{i}\vec{S}_{i}\cdot\vec{s}\delta(\vec{r}-\vec{R}_{i})$.
The exchange coupling arises due to the spin-spin interaction between spins in different atoms. Such an interaction is due to the superposition (overlap) between the wavefunction that describes conduction electrons on the surface
of a TI and bound electrons in the magnetic impurity on the surface. This is one of the most appealing from
the theoretical point of view because it not only gives a simple mechanism to develop the theory of the
topologically quantized magnetoelectric term, but it also allows us to look for unexpected effects that can
alter the magnetization dynamics \cite{garate and franz 2010,nomura and nagaosa 2010}. Therefore,
here we consider a $3D$ TI coated by thin film of layered two-dimensional ($2D$) Heisenberg ferromagnet. Indeed, magnetic materials (ferromagnetic or even antiferromagnetic) in two spatial dimensions may support topological excitations such as skyrmions and vortices. Particularly, vortices arise in classical magnetic systems containing an easy-plane anisotropy, which makes the spins to prefer to point along the $XY$-plane. For instance, easy-plane ferromagnets are described by the Hamiltonian $H=-J\sum_{i,j}[S_{i}^{x} S_{j}^{x}+ S_{i}^{y} S_{j}^{y}+ \lambda S_{i}^{z} S_{j}^{z}]$, where $J>0$ is the exchange constant, $0\leq \lambda < 1$ is the easy-plane anisotropy and $\vec{S}_{i}=(S_{i}^{x}, S_{i}^{y},S_{i}^{z})$ is the classical spin vector at site $i$. The spin field can be parametrized by two scalars $\Phi$ and $\varepsilon=\cos \Theta$, which are the azimuthal and polar angles in the unity spin sphere (internal space), as follows: $\vec{S}=\{\sqrt{1- \varepsilon^{2}} \cos\Phi, \sqrt{1-\varepsilon^{2}} \sin\Phi,\varepsilon\}$. Taken into account the most realistic cases of discrete lattices, and depending on the range of $\lambda$, such an easy-plane system supports two types of static vortices: the in-plane vortex (in which all spins are confined to the $XY$-plane\cite{???, cap-livro}) and the out-of-plane vortex (in which some spins around the vortex center can point perpendicularly to the $XY$-plane). Indeed, considering a critical value of $\lambda$ denoted by $\lambda_{c}$, then, for the range $\lambda <\lambda_{c}$, the stable excitation is the in-plane vortex while, for $\lambda >\lambda_{c}$, the out-of-plane vortex becomes stable \cite{Wysin94}. The stability of these solutions has only been determined via computer simulations. The critical anisotropy $\lambda_{c}$ depends on the lattice geometry: for the square lattice, $ \lambda_{c}=0.72$; similarly $\lambda_{c}=0.86$ for the hexagonal lattice and $\lambda_{c}=0.62$ for the triangular lattice \cite{Wysin94}. Qualitatively, similar results can be obtained for $2D$ easy-plane antiferromagnetic systems.

Our interest here is to investigate how an in-plane ferromagnetic vortex affects the electronic states lying on the surface of a $3D$ TI. Although $2D$ magnetic materials can support vortex excitations, there is, however, the experimental challenging of obtaining
and confining a single static vortex in these thin films \cite{Mertens89,Pires94,Hutchings86}. Even the
observation of these excitations in $2D$ layered easy-plane magnetic materials is not a simple task: Indeed, only indirect evidences for their appearance have been reported so far \cite{Mertens89,Zaspel95,Pereira96}. In addition, there is the
problem of getting a static single vortex living in the system for a long time \cite{Wysin96}. To overcome
these difficulties, a possibility should be the use of a vortex pinning mechanism. In $2D$ magnetic
materials, a nonmagnetic impurity (a vacancy) is able to pin the vortex center
\cite{Bogdan02,Pereira05,Paula04,Paula05,PereiraW06}, localizing the excitation around
it \cite{Bogdan02,Paula04,Paula05,Pereira03}. Another possibility (perhaps the simplest) should be through the use of
thin magnetic nanodisks over the TI. Really, depending on the size of a nanodisk, magnetic vortices are
naturally the ground state of these systems and they can be directly observed by experimental techniques\cite{Miltat02,Guslienko06}. Moreover, the vortex ground state is always out-of-plane (in the vortex core, the spins point perpendicularly to the disk plane) due to the competition between exchange and dipolar interactions. However, the presence of artificial defects intentionally inserted in the nanodisk (such as holes) \cite{Rahm03,Rahm04,Pereira05b,Pereira05c}, may destroy the usual out-of plane spin structure in the vortex core (after vortex capture), resulting in an essentially confined planar vortex centered at the hole
\cite{Silva08}.\\

Here we are interested in the in-plane vortex effects on the TI. Then, in principle, it can
be investigated either by applying a thin coating of magnetic material (with $\lambda <\lambda_{c}$ and
containing a small percentage of nonmagnetic impurities) on the TI or by putting a magnetic nanodisk (with a
hole, preferentially in the disk center) on the top of the TI.  In the context of the first situation,
recently, the authors of Ref.[\onlinecite{nomura and nagaosa 2010}] have theoretically studied the magnetic
textures, such as domain walls and vortices in a ferromagnetic thin film deposited on the surface of a
TI. These authors have found that the magnetic textures interacting with the topological surface states are
electrically charged. The experimental observation of massive Dirac Fermion on the surface of a
magnetically doped TI was reported in Ref.[\onlinecite{y. l. chen}], where the authors introduced
magnetic dopants into three-dimensional TI $Bi_{2}Se_{3}$  and observed the massive Dirac fermion state by
angle-resolved photoemission. Other aspects as hedgehog spin texture\cite{xu}, magnetic scattering
of the carriers on the surface\cite{zazunov}, magnetic ordering \cite{abanin, tome} and
topological magneto-electric effect \cite{Essim, Qi, jakson-cone} were also
investigated.

Considering an in-plane closed flux magnetized state interacting with a $3D$ TI, we have realized that the interaction between the surface carriers and the vortex leads to bound-states and spin polarized zero-modes, besides scattered states. Such a framework provides a mechanism to obtain spin polarized current based on the zero-mode states; conversely, such a physical mechanism appears to be useful for experimental magnetism to precise and efficiently detection of magnetic vortices in layered magnetic thin films. Our article is outline as follows: in Section $II$ we present the model which takes in to account the basic properties of the TI surface electronic states along with their coupling to the magnetic moments (spins) lying on the ferromagnetic layer. Section $III$ is devoted to the zero-mode states induced by the vortex. They are spin polarized modes that form a fully polarized current in the TI around the center of the magnetic vortex. Additionally to its basic features, we also point out possible utilities for such a special current. In Section $IV$, the remaining bound-states are presented and discussed, completing the `atomic picture' provided by the compound system vortex-TI charge carries. We finally close our paper by pointing out our conclusions and prospects for future investigation.

%%%%%%%%%%%%%%%%%%%%%%%%%%%%%%%%%%%%%%%%%%%%%%%%%%%%%%%%%%%%%%%%%%%%%%%%%%%%%%%%%%%%
\section{The model}

The low energy excitations on the surface of a TI are described by a Dirac-like equation in a
two-dimensional plane. When the chemical potential is confined between valence and conduction
bands (in the ground state), the positive energy modes (that describe free conduction electrons)
are empty, while the negative energy modes (that describe bound valence electrons) are filled
with valence electrons. The mass term in the equation expresses the gap of energy between the
conduction and valences bands (a trivial insulator) and a vanishing mass term leads to a gapless
spectrum conductor or TI surface modes\cite{jackiw-arxiv-2011}. The latter ones offer a unique platform to investigate the physics of robust Dirac points.\\

Here, we analyze the simplest case of a single Dirac point (for instance,
$Bi_{2}Se_{3}$, $Bi_{2}Te_{3}$ and
$Sb_{2}Te_{3}$ \cite{zhang-nature}). Let us consider the surface of the $3D$ TI in the $XY$-plane (its bulk extends to $z<0$ region) coated by a ferromagnetic thin film supporting an in-plane vortex. Therefore, the low-energy effective Hamiltonian for the TI
surface states coupled to the vortex magnetic moments (spins) reads:
\begin{equation}
H=H_0 + H_{int}
\end{equation}
where
\begin{equation}\label{H-free}
H_0= v_F\psi^\dag(\vec r)(k_x\sigma^x + k_y\sigma^y)\psi(\vec r)\,,
\end{equation}
and
\begin{equation}\label{H-inter}
H_{int}=-\sum_{i=x,y,z}\Delta_i \vec M_i(\vec r)\cdot \vec s(\vec r)\,.
\end{equation}

Above, $\vec s(\vec r)=\psi^\dag(\vec r) \vec\sigma \psi(\vec r)$ is the spin density of
surface electrons in the position $\vec r$, $\vec M(\vec r)$ is the magnetic moment applied on the surface of the TI, which
characterizes the magnetic configuration on the surface and $\Delta_i$ is the ferromagnetic exchange constant \cite{liu-prl, TI review}, accounting for the coupling between the electrons lying on the TI surface and the magnetic moments composing the magnetic material (exchange-type proximity effect; note that $H_{int}$ is minimized when the electrons spins align to the dipoles). For instance, regarding the material $Sb_{2}Te_{3}$ doped with vanadium, the surface
exchange parameters $\Delta_x,\,\Delta_y,\,\Delta_z$ are estimated to be of the order of $0.1-0.5 eV$, relying on the
overlap between the wave-functions \cite{liu-prl}. For the case of a ferromagnetic ordered state perpendicular to the surface, the interaction Hamiltonian (\ref{H-inter}) has the form of a mass term for the surface electrons proportional to $\sigma_z$, being the carriers massive with a dispersion relation like:
\be
E=\pm v_F\sqrt{k_x^2+k_y^2+(\mu v_F)^2}\,,
\ee
where the positive and negative signs describe the conduction electrons and the valence
electrons (or holes), respectively. Whenever this is in order, a gap opens, $E_g= 2\mu v_f^2$, rendering TI to be a quantum Hall insulator. As we shall show below, the magnetic vortex also breaks TRS in the surface, but does not induces a mass gap, like above. As consequences we shall find electronic bound-states and zero-energy modes around the vortex.

%%%%%%%%%%%%%%%%%%%%%%%%%%%%%%%%%%%%%%%%%%%%%%%%%%%%%%%%%%%%%%%%%%%%%%%%%%%%%%%%%%%%%%%%%%%%%%%%%%%

\subsection{Magnetic vortex configuration}
\begin{center}
\begin{figure}[!h]

\includegraphics[angle=0.0,width=8cm]{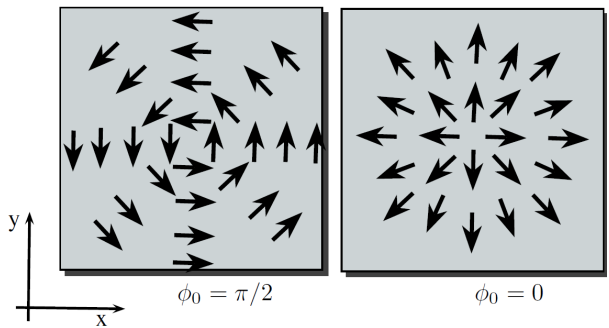}
\caption{(Color online) Examples of ferromagnetic vortex-like patterns with unity charge, $q=+1$, and with profiles $\Phi_0=+\pi/2$ (curly-type) and $\Phi_0=0$ (radial-like). The depicted arrows represent the magnetization vector field, $\vec{M}$.} \label{fig1-vortex}
\end{figure}
\end{center}

An in-plane vortex in the continuum limit may be described by\cite{Mertens89}:
\begin{equation}
\varepsilon=\varepsilon_{v}=0, \quad
\Phi=\Phi_{v}=q\arctan\left[\frac{y-y_{0}}{x-x_{0}}\right] + \Phi_{0},
\end{equation}
where $q$ accounts for its topological charge (vorticity), centralized at $(x_{0},y_{0})$, while $\Phi_{0}$ is a constant determining its profile (in Fig.\ref{fig1-vortex} we depict two examples of unity charge vortex, $q=1$, with distinct profiles). For such a planar vortex centered at the origin $(x_0,y_0=0,0)$, so that $\Phi_v= \theta +\Phi_0$, its magnetization, $\vec{M}= M(\sqrt{1- \varepsilon_v^{2}} \cos\Phi_v, \sqrt{1-\varepsilon_v^{2}} \sin\Phi_v,\varepsilon_v)$, in 2D polar coordinates $(r,\theta)$, may be written as $\vec{M} (\vec{r})=  M_r \hat{r}+ M_\theta \hat{\theta}$ with $M^{2}=|\vec{M}|^2 = constant$ (with $M \equiv|\Delta_iM_i|$ (hereafter we normalize $\Delta_i\equiv 1, \,\forall i$)). Note that $M_r=M\cos(\Phi_0)$ and $M_\theta=M\sin(\Phi_0)$ account for the radial- and curly-type profiles of the vortex, so that for $\Phi_0=0$ one gets $(M_r, M_\theta)=(M,0)$ a radial vortex while for $\Phi_0=\pm /pi/2$ a curly-type vortex with $(M_r, M_\theta)=(0,\pm M)$ is observed (see Fig. \ref{fig1-vortex}). Therefore, the interaction Hamiltonian
(\ref{H-inter}) becomes:

\begin{equation}
H_{int} = - \psi^\dag(\vec r) (M_r \hat{r} +M_\theta\hat{\theta}) \cdot\vec\sigma\psi(\vec r)\,.
\end{equation}
This term does not open a mass gap in the spectrum, like an ordered ferromagnet does; It rather induces the appearance of bound-states, including zero-modes, in the electronic spectrum of the composed system, as shown below. Thus, the dynamic of the low energy states is
described, in polar coordinates $(r,\,\theta)$, by the Hamiltonian:

\begin{eqnarray}
\label{hamiltoniana matriz}
&&{\cal H} = -i\hbar v_F\times \nonumber \\
&&\left(\begin{array}{cc}
       0 & e^{-i\theta}\,\big(\,\frac{\partial}{\partial r} - \frac{i}{r}\frac{\partial}{\partial\theta} -
       \frac{M_\theta-iM_r}{\hbar v_F}\,\big)\\
       e^{i \theta}\,\big(\,\frac{\partial}{\partial r} + \frac{i}{r}\frac{\partial}{\partial\theta} +
       \frac{M_\theta +iM_r}{\hbar v_F}\,\big) & 0
      \end{array}\right)\,.
\end{eqnarray}

Hence, the eigenvalue problem ${\cal H}\psi=E\psi$ can be solved analytically by the usual ansatz:
\be
\label{symmetric spinor}
\psi( \vec r)= \left(\begin{array}{c}
        e^{im \theta}\varphi(r) \\
        e^{i(m+1)\theta}\chi(r)
      \end{array}\right)\,, \quad m=0,\,\pm1,\,\pm2,\,\,\ldots\,,
\ee
where the components of the spinor $\varphi(r)$ and $\chi(r)$ satisfy the first order
coupled differential equations given by:
\be
\label{g e f}
 \bigg( \frac{d}{dr} + \frac{m+1}{r} - \frac{M_\theta-iM_r}{\hbar v_F} \bigg) \chi(r)
= i\frac{E}{\hbar v_F} \varphi(r)\,,
\ee
\be
\label{f e g}
 \bigg( \frac{d}{dr} - \frac{m}{r} + \frac{M_\theta+iM_r}{\hbar v_F} \bigg) \varphi(r)
= i\frac{E}{\hbar v_F} \chi(r)\,,
\ee
Before to proceed further, it is noteworthy to remark that $\pm iM_r$ terms in equations above work like a {\em centrifugal potential}, drifting electronic states and eventually preventing them of getting bound to the vortex center region. Actually, if the vortex is purely radial, $\Phi_0=0$, no normalized solutions for eqs. (\ref{g e f})-(\ref{f e g}) with $E\leq0$ is supported in the physical spectrum; in this case, only scattered states are observed. In addition, if the vortex has both radial- and curly-type components, even a tiny $M_r$ still rules out the bound-states, $E< |\vec M|$; the remaining zero-modes solutions oscillate in radial and angular directions as well, but also decrease their amplitudes along $r$, what may be clearly realized by setting $M\to M_\theta \pm i M_r$ in expressions (\ref{13})-(\ref{14}), eventually rendering such modes to be not fully polarized if $M_r\neq0$. Therefore, for explicitly working out zero-modes and bound-states solutions we consider only the curly-type vortex, say, hereafter we set $\Phi_0=\pm \pi/2$ so that $\vec{M}=\pm M_\theta\hat{\theta}\equiv \pm M \hat{\theta}$, with $M>0$, what appears to be the most interesting scenario.\\

%%%%%%%%%%%%%%%%%%%%%%%%%%%%%%%%%%%%%%%%%%%%%%%%%%%%%%%%%%%%%%%%%%%%%%%%%%%%%%%%%%%%%%%%%%%%%%%%%%
\section{Zero-energy modes}

Firstly, we investigate the zero energy solutions (zero-modes) of the equations (\ref{g e f}) and (\ref{f e g}) for the case of a planar vortex with $\phi_{0}=\pi/2$. When the Dirac equation has a homogeneous position-independent mass term, the solutions are
as usual: continuum solutions with positive ($E>0$) and negative ($E<0$) energies. However, in
the presence of a topologically non-trivial background or defect, a mass term position
dependent can appear and the spectrum contains (in addition to the habitual positive and negative
energy modes) also isolated zero-energy mode states. The zero-energy solutions of
the Dirac equation do not require a specific form for the inhomogeneous mass profile; instead,
they only require that the mass profile belongs to a nontrivial topological class
or, when coupled to a magnetic vortex, a nontrivial vorticity. Then,
the presence of zero-energy modes can be established, a priori, by mathematical index theorems,
which relate the occurrence of these zero modes to the geometry and topology of the space on which the
Dirac equation is stated \cite{jackiw-arxiv-2011,Jackiw84,Ansourian77}. The index theorem offers an analytic
tool that relates the zero modes of elliptic operators (like the Dirac Hamiltonian in $2+1$ dimensions)
with the geometry of the manifold on which these operators are defined \cite{index}.
This theorem can be employed to obtain information about the low energy behavior
of the TI and, in particular, about its conductivity properties. A general study of the
index theorem in TI will be presented elsewhere \cite{jakson-index}.

\begin{center}
\begin{figure}[!h]
\includegraphics[angle=0.0,width=8cm]{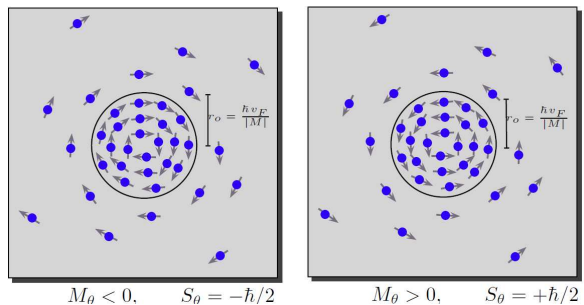}
\caption{(Color online) By breaking time reversal symmetry a ferromagnetic vortex forces the electrons spins, confined to the topological insulator surface, to align along its magnetization direction, according to Eqs. (\ref{zero-energy1}) and (\ref{zero-energy2}) . Such an alignment is fully accomplished by zero-mode fermions that appear concentrated in a region of radius $r_0$ around the vortex center, what yields the possibility of highly spin-polarized zero-mode current.} \label{zero-mode-fig}
\end{figure}
\end{center}

Now, we investigate the possibility of occurrence of zero-modes on a surface of the
TI without a mass profile in the Dirac equation due to the presence of a non-trivial background
for the surface carriers. We study this occurrence based on analytical procedures
by explicitly solving the Dirac equation. In the case of zero energy solutions, the radial
components of the spinors (\ref{g e f}) and (\ref{f e g}) must satisfy:
\be \label{13}
\bigg(\frac{d}{dr}-\frac{m}{r}+\frac{M}{\hbar v_F}\bigg)\varphi(r)=0
\ee
\be \label{14}
\bigg(\frac{d}{dr}+\frac{m+1}{r}-\frac{M}{\hbar v_F}\bigg)\chi(r)=0.
\ee 
There are two possibilities, depending on the orientation of the local magnetization: when the magnetization is oriented clockwise, $\vec{M_\theta} = -M$, the physical zero-modes read:
\be\label{zero-energy1}
\psi( \vec r)=a_m
\left(\begin{array}{c}
        0 \\
        1
      \end{array}\right)
      \frac{e^{-\frac{M}{\hbar v_F}r}}{r^{m+1}}e^{i(m+1)\theta}\,,
\ee
where $a_m$ are the normalized amplitudes given by
$a_m=\frac{1}{\sqrt{2\pi(|2m+1|!)}} \bigg(\frac{2M}{\hbar v_F}\bigg)^{|m+1/2|+1/2}$. Solution
(\ref{zero-energy1}) can be normalized only if $2m+1<0$; in addition, these solutions are spin-polarized with
$s_\theta=-\frac\hbar2$, so that the vortex provides a mechanism for supporting spin-polarized zero-energy current on the TI surface (see Fig. \ref{zero-mode-fig}). On the other hand, if the local magnetization is oriented counterclockwise, $M_\theta= -M$, the zero-mode solutions are spin polarized with $s_{\theta}=+\frac\hbar2$ and given by:
\be\label{zero-energy2}
\psi( \vec r)=b_m
\left(\begin{array}{c}
        1\\
        0
      \end{array}\right)
r^m e^{-\frac{M}{\hbar v_F}r} e^{im\theta}\,,
\ee
with $b_m=\frac{1}{\sqrt{2\pi(2m+1!)}} \bigg(\frac{2M}{\hbar v_F}\bigg)^{m+1}$. This solution has physical
meaning only if $2m+1>0$. Both sets of spin-polarized zero-modes carry integer electronic charge:
\be
Q/e^-=\int d^2r\,j^0=\int d^2r\psi^\dag\psi=1\,.
\ee
Physically, it is precisely the coupling between the orbital angular
momentum of the carriers and the magnetic moments of the vortex that yields polarized charged zero-modes confined to a circular region with radius $r_0\equiv\hbar v_{F}/M$, around the vortex center. Taking $Sb_2Te_2$ as an example, one has $\hbar v_F\approx 3.7 {\rm eV\, \AA}$ while $M$ is typically $\sim 50 {\rm meV}$ giving us $r_0 \approx 74 {\rm \AA}$ (see Fig.\ref{zero-mode-fig}). It is important to remark that, in this case, there is only one fermion per quantum state (which is denoted by the quantum number $m$); indeed, as we have seen, the vortex chirality requires that these states must be polarized. Therefore, in principle, such zero modes may contain a very large number of polarized carriers (one per each possible $m$) for a given vortex chirality.
This very large number of polarized carriers can be used to establish a full spin polarized current over the TI surface or, more specifically, the coupling vortex-TI can be used as a mechanism for construct electronic devices able to induce spin polarized currents. On the other hand, from the magnetism point of view, the observation of this special current over the TI could be an alternative technique for indirectly detecting a vortex and its chirality in magnetic thin films.

%%%%%%%%%%%%%%%%%%%%%%%%%%%%%%%%%%%%%%%%%%%%%%%

%%%%%%%%%%%%%%%%%%%%%%%%%%%%%%%%%%%%%%%%%%%%%%%%%%%%%%%%%%%%%%%%%%%%%%%%%%%%%%%%%%%%%%%%%%%%%%%%%%%
\section{Bound states}

In addition to the zero-modes, there also exist additional bound-state solutions of Eqs. (\ref{g e
f}) and (\ref{f e g}) with $E<|\vec M|$. The solutions correspond to internal modes in which the electrons wave-functions undergo harmonically oscillations around the vortex center. Bound-states can be induced in $2D$ or $3D$ TI's by vacancies or holes in the band gap. These defects can affect the transport properties of TI's via additional bound-states near the system boundary due to
the coexistence of in-gap bound-states near the edge or surface states\cite{wen bound-states}. Here, this effect arises due to the exchange coupling between the spin of the surface carriers and the magnetic moment of the vortex in the surface.

To obtain the bound-states we proceed in the usual way by looking for solutions of  Eqs.
(\ref{g e f}) and (\ref{f e g}) with energy, $E<|\vec M|$. By Decoupling these equations, we obtain the following second order differential equation for the up component of the spinor (down component satisfies a similar expression, as below):
\begin{eqnarray}
\label{edo2ordem}
\frac{d^2 \varphi(r)}{dr^2}+\frac1r\frac{d\varphi(r)}{dr} &+& \nonumber \\
\bigg(- \frac{m^2}{r^2}+\frac{M}{\hbar v_F}\frac{2m+1}{r}&+&\frac{E^2-M^2}{\hbar^2v_F^2}\bigg)\varphi(r)=0\,.
\end{eqnarray}
To solve this equation, we use the correspondence between this problem and the $2D$ quantum Kepler problem ($2D$ hydrogen atom). Such an equation admits solutions for electrons restricted to move in a plane around the nucleus, due to the potential $1/r$ with \cite{H2} $r^{2}= x^{2}+y^{2}$. The energy eigenvalues for the bound-states read:
\be
\label{autovalores}
E(n\,,\,m) =\pm |\vec M|\Bigg[1-\frac{(m+1/2)^2}{(n+|m|+1/2)^2}\Bigg]^{1/2}\,,
\ee
with $n\,=\,1\,,\,2\,,\,3\,,\,\ldots$ and $|m|=0\,,\,1\,,\,2\,,\ldots\,n-1$. The radial
eigenfunction can be obtained in terms of the hypergeometric function \cite{H2}:
\begin{eqnarray}
\varphi_{n,m}(r)&=&\frac{\alpha_{n,m}}{(2|m|)!}
\bigg[ \frac{(n+|m|-1)!}{(2n-1)(n-|m|-1)!} \bigg]^{1/2}\times \nonumber \\
&&(\alpha_{n,m} r)^{|m|}e^{-\alpha_n r/2}\times \nonumber \\
&&_1F_1(-n+|m|+1,\,2|m|+1,\,\alpha_{n,m} r)\,.
\end{eqnarray}
Here, $\alpha_{n,m}=\frac{M(2m+1)}{(n-1/2)\hbar v_F}$ and $_1F_1$ is a
confluent hypergeometric function. The component $\chi_{n,m}(r)$ of the spinor can be obtained
from $\varphi_{n,m}(r)$ by using equation (\ref{g e f}).
\begin{center}
\begin{figure}[!h]
\includegraphics[angle=0.0,width=8cm]{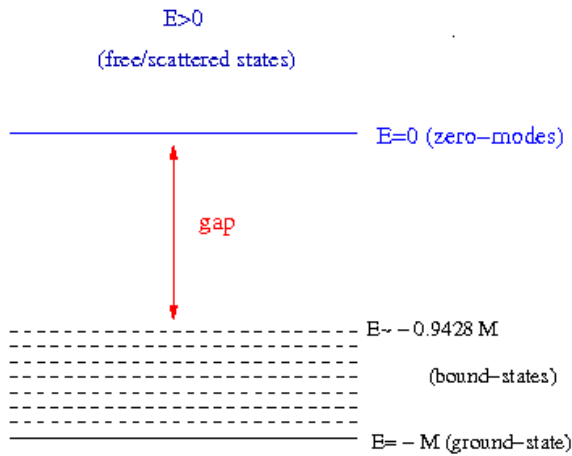}
\caption{(Color online) The `atomic' energy levels allowed for zero-modes and bound-states 
(only $E<0$ bound states are shown) in the composed vortex-topological insulator system. An almost continuum spectrum is allowed between $-M$ and $\approx -0.94M$, where the most energetic bound-state occurs, while $\delta\approx +0.94M$ gaps this state from zero-modes. The bound states are symmetric in relation to the zero energy states, having positive energy bound states.}
\label{fig2-energy-scale}
\end{figure}
\end{center}

Note that the infinite number of energy levels goes from $-M$ (the ground state) to 
$-2\sqrt2M/3\approx-0.9428M$ (the most excited state with $E<0$) but, in contrast to the $2D$ Hydrogen atom, 
the ground state is characterized by $n=\infty$ (and the negative signal in (\ref{autovalores})) 
while the most excited state is characterized by $n=\infty$ (and the positive signal in (\ref{autovalores})) (Fig. 
\ref{fig2-energy-scale} illustrates the energy levels for such an `atom'). Here, contrary to Hydrogen atom
there are bound states with
positive energy, and goes from 
$2\sqrt2M/3\approx0.9428M$ to $M$.
The Bohr radius is $a_0=\frac{2\hbar v_F}{M}$, which is on the order of $148\AA$ for typical values of 
$v_{F}$ and $M$. The most probable radius for these bound-states are $(n+1/2)a_0$ (let us recall that $a_0$ 
is much smaller than the size of typical magnetic nanodisks containing a vortex ground state. This is 
particularly important if these nanoscaled systems are employed for TI coating).  In addition, the most 
excited bound-state radial wavefunctions (with $E<0$) are:
\be
\label{varphi10}
\varphi_{10}(r)=\frac{2M}{\hbar v_F}e^{-Mr/\hbar v_F}\,,
\ee
\be
\label{chi10}
\chi_{10}(r)=\pm\frac{3i}{\sqrt{2}}\bigg(\frac{2M}{\hbar v_F}+\frac{m}{r}\Bigg)e^{-Mr/\hbar v_F}\,,
\ee
from what becomes more evident that the effective potential brought about by the magnetic vortex is able to confine
the TI surface electrons in a limited region around the vortex center. Similarly to the zero energy modes, the
region of larger probability is a circle with a radius on the order of $\hbar v_{F}/M$, depending only on the
value of the local magnetization $M$ (decreasing as $M$ gets higher). In addition, bound-states with normalized wavefunctions are possible only if $M(2m+1)>0$. Therefore, there are two possibilities: if $M>0$ then $m >-1/2$, while for $M<0$,  the values $m <-1/2$ must be taken. As a consequence, all bound electrons move around the same sense, determined by the vortex magnetization chirality (see Fig. \ref{zero-mode-fig}). Although all bound electrons move around the same sense, there are different probabilities
associated with each spin mode (see (\ref{varphi10}) and (\ref{chi10}) for the most excited state, for example); then the current bound around the vortex center has a partial polarization.\\

Besides the bound-states, a continuum spectrum of scattered states should also take place; they could be parametrized by a continuum wave-vector $\vec{k}$ whose associated wave-functions solutions must be obtained from equation (\ref{edo2ordem}). For these states, the relevant results can be obtained in terms of the phase-shifts of the carriers, cross-section and resistivity, and will be presented elsewhere \cite{jakson2}.

Some additional remarks are now in order. In general, the Dirac equation has negative and positive energy solutions. The negative energy solutions correspond to the states in the valence band while the positive energy solutions correspond to the states in the conduction band (considering the chemical potential located at $E=0$). In the ground-state, all the negative energy levels are filled and, therefore, the ground-state is degenerate (since the zero modes energy are degenerate). Then, what is the ground-state at TI surface state in the presence of a magnetic vortex? In other words, how could one treat the zero-energy modes whenever constructing the ground-state of the TI-ferromagnetic composed material? Should the states be empty like the case of positive energy or should they be filled as they are for negative energy states?

If they are filled (or partially filled) when an electric current is established over the TI with a vortex (and
only zero-modes are excited), then the current is spin polarized, being the polarization given by sense of
the magnetization. However if the bound-states are excited to the conduction band (overcoming $\delta=2\sqrt{2}M/3\approx 0.9428M$), then the current has a gap and the current will display steps when increasing the external electrical field similar to the Hall effect under increasing of the external magnetic field.\\

%Another point that deserves attention is the effect of the vortex profile on the electrons. If our anterior analyze were performed for a planar radial vortex configuration in which the magnetization develops a radial component (i.e., with the constant $\phi_{0}$ assuming any possible value different from $\pi/2$), then, it is easily to see that the zero-modes are not spin polarized any more while all bound-states, with $E<0$, are ruled out. Indeed, even a very small radial component for the spins forming a vortex is able to generate a term in the Hamiltonian which makes the solutions with negative energy to be unstable; they can not be normalized. Thus, there is a kind of centrifugal effect as a result of the spins radial component (even for minimal radial component), which expel the electronic states (with $E<0$) from the vortex center. Therefore, although the entirely possible values of the constant $\phi_{0}$ can give vortices with different configurations, only the case $\phi_{0}=\pi/2$ is relevant for developing the states described here.\\

It is also noteworthy to mention that the authors of Ref. [\onlinecite{Katsnelson gauge fields TI}] have treated a similar problem of a magnetic vortex coating the surface of a TI. However, in this work, the authors have considered the question of whether
an effective gauge field would be induced by strain on the surface of a topological insulator. They have shown that the gauge fields arise when the surface is coated with an easy-plane ferromagnet with magnetization parallel to the surface \cite{Katsnelson gauge fields TI}.

%
%%%%%%%%%%%%%%%%%%%%%%%%%%%%%%%%%%%%%%%%%%%%%%%%%%%%%%%%%%%%%%%%%%%%%%%%%%%%%%%%%%%%%%%%%%%%%%%%%%%%
%\section{Local density of States}
%
%The local density od states (LDOS) can be obtained from the corresponding retarded Gren's function
%as
%
%\be
%N(\vec r\,,\,\omega)=-\frac2\pi{\rm Im\,Tr\,}G^R(\vec r\,,\,\vec r\,;\,\omega)\,,
%\ee
%%
%being the factor $2$ due to spin degeneracy. For a flat topological insulator without
%impurity one gets the (zero order) Green?s function:

%%%%%%%%%%%%%%%%%%%%%%%%%%%%%%%%%%%%%%%%%%%%%%%%%%%%%%%%%%%%%%%%%%%%%%%%%%%%%%%%%%%%%%%%%%%%%%%%%%%
\section{Discussions and conclusions}

In summary, we have investigated the effects of in-plane vortices on the surface of a $3D$ strong topological insulator with one Dirac cone. Only a closed flux magnetization (a vortex with $\Phi_{0}=\pi/2$) are important for our conclusions. Indeed, we have shown that the interplay between topologically protected surface states and TRS breaking by an in-plane magnetic vortex (in which $\Phi_{0}=\pi/2$) coating TI surface does not generate a mass gap; It rather yields bound-states and localized spin polarized charged zero-modes. Although zero-modes have already predicted in a somewhat similar field theoretical scenario\cite{Jackiw84,Ansourian77}, our work has not only re-accessed the problem, but has also the merit of taking it to physical realization with current technology. Considering the bound-states, the very presence of the vortex generates an effective potential for the TI surface carriers which is equivalent to the problem of a two-dimensional hydrogen atom. The range of energy levels of the bound-states is very short and goes from $-\mid M \mid$ to $-0.9428 M$ (and $0.9428 M$ to $|M|$) (comprising infinite possible states in this range), which is energetically separated from the zero-modes, $E=0$, by a large gap $\delta \approx 0.9428M$; a picture of such energetics may be seen in Fig. \ref{fig2-energy-scale}). Zero-modes and bound-states decay exponentially as $\exp(- Mr/\hbar v_{F})$ away from the vortex center, which means that some electrons are essentially localized in a circular region with radius $r_0=\hbar v_{F}/M$ (around $70 {\rm \AA}$ for typical TI's with ``atomic'' binding energy $\sim 50 {\rm meV}$. Physical systems supporting stable zero-modes deserve special attention, for instance, in topological quantum computation for the ground-state degeneracy remains unaffected by small perturbations, what can render error free quantum information encoding.\\

Instead of large magnetic films, nanoscale magnets can also be used to form the composed system with TI's. In this case, special care must be taken because vortices in confined structures must have out-of-plane components near its center, what opens up a gap in the TI electronic states. The remaining effects of a vortex, as studied here, should remain qualitatively unaffected, still supporting zero-modes and bound-states on the TI surface. Even in the case of nanomagnets, in-plane vortices may take place, for instance, if a cavity is intentionally inserted at the center of the disk\cite{Rahm03,Rahm04} or we employ other geometries, such as a nanotorus \cite{C-Santos10}. Another interesting possibility is to coat TI surface with an elliptic thin disc supporting a pair vortex-antivortex, with opposite chiralities. On the TI surface it would be induced two distinct zero-modes fermionic current flowing in opposite directions, around each of the vortex center.

Finally, it should be remarked that besides of being of potential utility for novel magneto-electric mechanisms in spintronics and related branches, such special modes can be faced as an alternative technique for detecting magnetic vortices in magnetic thin films and in nanomagnets as well.\\

\vskip .5cm
\centerline{\bf Acknowledgments\\}
The authors thank O.M. Del Cima, A.H. Gomes and D.H.T. Franco for useful discussions. They are also grateful to CNPq, FAPEMIG and CAPES (Brazilian agencies) for financial support.

\end{document}